\documentclass[aps,prl,twocolumn,amssymb,groupedaddress,floatfix]{revtex4}
\usepackage{graphicx}

\begin{document}


\title{The Geometry of Crumpled Paper}
\author{Daniel L. Blair\footnote{Present Address: Department of
    Physics $\&$ DEAS, Harvard University, Cambridge, Massachusetts
    02138, USA } and Arshad Kudrolli} 
\affiliation{Department of Physics, Clark University, Worcester,
  Massachusetts 01610} 
\date{\today}

\begin{abstract}
We measure the geometry of a crumpled sheet of paper with laser-aided
topography and discuss its statistical properties. The curvature of an
elasto-plastic fold scales linearly with applied force. The curvature
distribution follows an exponential form with regions of high
curvature localized along ridges. The measured ridge length
distribution is consistent with a hierarchical model for ridge
breaking during crumpling. A large fraction of the ridges are observed
to terminate without bifurcating and the ridge network connectedness
is not as complete as anticipated. The self-affinity of the surface is
characterized by a Hurst exponent of $0.72\pm 0.01$ in contrast with
previous results. 

\end{abstract}



\maketitle



A crumpled piece of paper is an interesting and ubiquitous example of
a stress induced morphological transformation in thin sheets. As a
sheet is deformed, the bending energy becomes localized and the
resulting crumpled shape is often thought of as a network of connected
line-like ridges. These characteristic topologies are found in a
variety of objects  ranging from biological systems to engineering
applications -- for example, polymerized vesicle membranes and crumple
zones in automobile bodies. In all situations, the response of the
material can vary from purely elastic deformations to elasto-plastic
depending on the applied stress. Much of the theory for crumpled
membranes and shells focus on either equilibrium transformations found
in biology~\cite{nelson87}, or the elastic properties of macroscopic
sheets~\cite{kramer_prl97,cerda_prl03}.  A number of recent
studies~\cite{cerda_prl98,cerda_n99,arezki_n00,cerda_n02}, treat the
the geometry of developable cones which form the geometrical
foundation for structures found when elastic sheets are subjected to
point deformations.
However, few studies exist that directly explore the crumpled state
when elasto-plastic deformations occur.

Early experiments focused on the fractal dimension of the ball that
circumscribes a flat sheet once crumpled~\cite{gomes89}.  The acoustic
emission from sheets that have been crumpled and unfolded display
power-law scaling in the distribution of click
energies~\cite{alex_pre96,sethna}.  By postulating that the ridge
lengths are proportional to the energy released in acoustic emission,
the authors indirectly measured the ridge distribution. However, very
few experiments which directly access the surface leave many
fundamental questions unanswered.

In this paper, we report new experiments designed to directly measure
the geometry of crumpled paper using non-invasive, laser-aided
topographical reconstruction. Large sheets of paper are crumpled to a
ball of fixed radius and then unfolded to reveal the plastic
deformations made by the crumpling process. The surfaces are scanned
by a laser sheet and the coordinates of the surfaces are
reconstructed. Thus, we directly access the local curvature of the
sheet and measure the properties of the ridges and their structure.
We also examine the curvature of a single fold as a function of
applied force to probe the elastic and plastic deformation which occur
during crumpling.

\begin{figure}
(a) \includegraphics[width=0.7\linewidth]{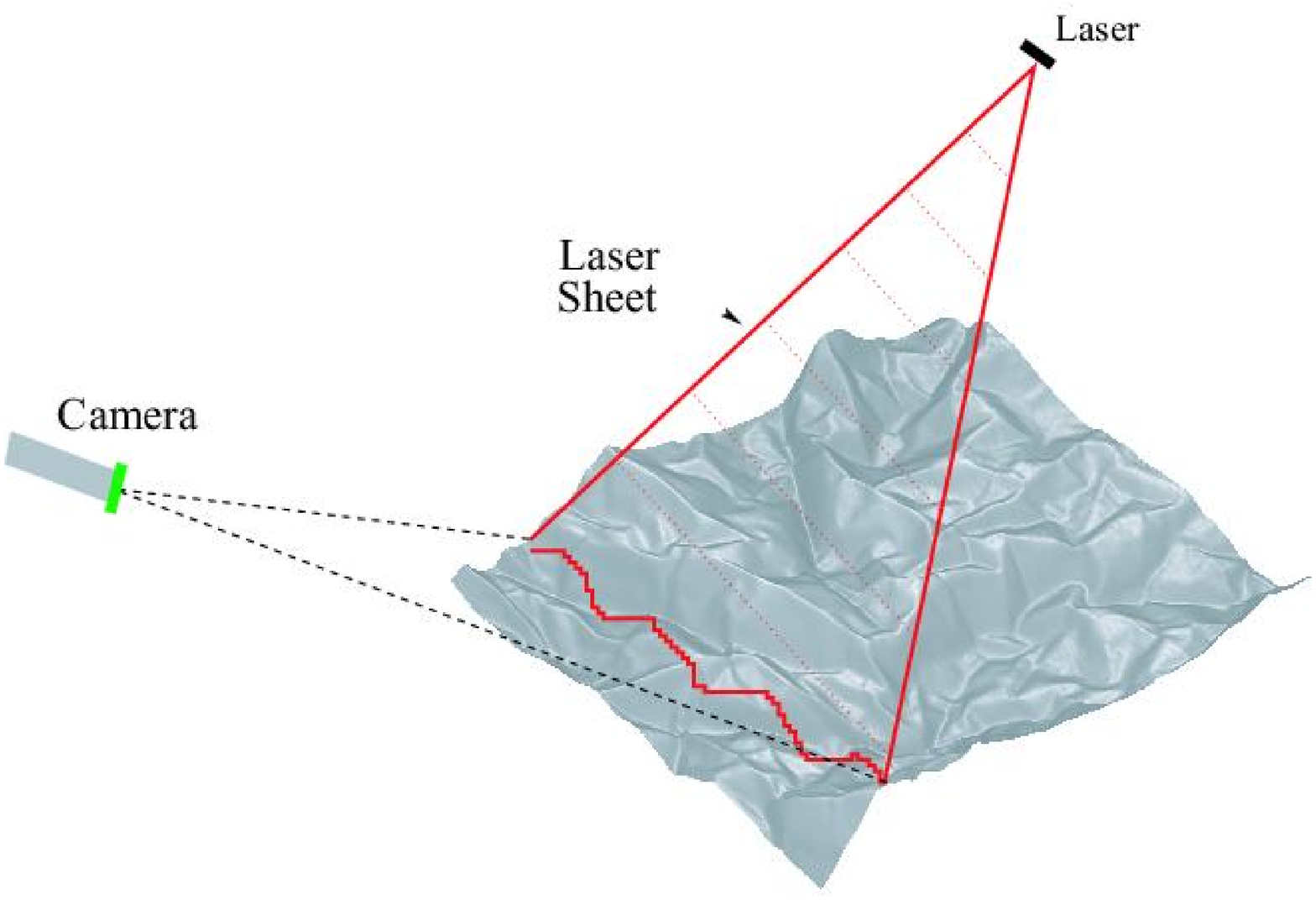}\\ (b)
\includegraphics[width=0.7\linewidth]{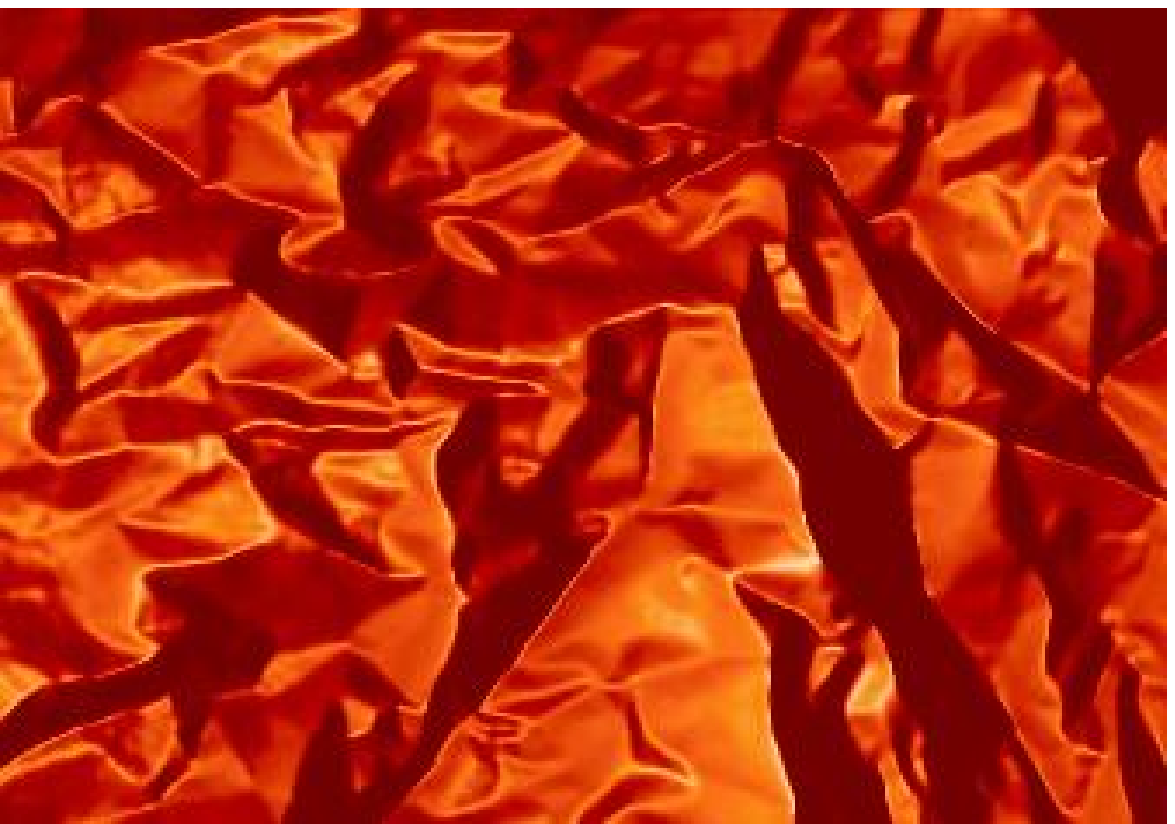}
\caption{(a) Schematic diagram of the experimental apparatus. (b)
  Example of a crumpled paper measured with laser-aided topography and
  rendered with shadow lighting.}
\label{fig:paper_schem}
\end{figure}

A schematic of the experimental configuration is shown in
Fig.~\ref{fig:paper_schem}(a). Three kinds of paper are used (see
Table~\ref{table:thick}.) The paper are first crumpled to ``spheres''
of radius $R_c \approx 120$ mm and then unfolded to a partially flat
state.  The uncrumpling process consists of opening the sphere, taking
care not to tear the sheet while also making sure that we do not
flatten the scars that are left on the paper.  The opened sheets are
then placed into the scanning setup for measurement. A laser with a
cylindrical lens produces a sheet of light approximately in the
vertical direction and a CCD camera with a resolution of $1024 \times
768$ pixels is placed at an angle to image the intersection of the
laser light with  the surface. When imaged from the side, the bright
points where the laser light intersects the surface is proportional to
its height [see Fig.~\ref{fig:paper_schem}(a)]. Each column of the
image is first scanned for the highest pixel value and that element is
used for the brightness weighted centroid along that particular
column. Care is taken to ensure that the camera is mounted so that the
laser light does not ``disappear" behind the structure that is being
scanned. The laser sheet is rotated with a stepper motor which allows
the entire surface to be scanned. After appropriate scaling and
calibration, the crumpled surface is measured to within the thickness
of the paper (see~\cite{blair04} for further details.) An example of a
crumpled surface is shown in Fig.~\ref{fig:paper_schem}(b), and
corresponds to a 450 mm $\times$ 588 mm sheet of paper.  The resulting
``surface'' is a matrix of elements whose value defines a height.
Morphological features are extracted using a local quadratic fit to a
local fitting region~\cite{wood}.

\begin{table}[t]
\centering
\begin{tabular*}{0.85\linewidth}{@{\extracolsep{\fill}} c c c }
\hline Mass (g mm$^{-2}$ ) & $h_p$ (mm) & Size (mm $\times$ mm) \\
 \hline \hline   1.16 x$10^{-6}$ & 0.156 & $560 \times 610$ \\    1.89
 x$10^{-6}$ & 0.207 & $560 \times 610$ \\ 2.97 x$10^{-6}$ & 0.360 &
 $430 \times 355$ \\ \hline
\end{tabular*}
\caption{Mass, thickness, and size of paper used in the crumpling
  experiments.}
\label{table:thick}
\end{table}

\begin{figure}
\begin{center}
\includegraphics[width=0.35\textwidth]{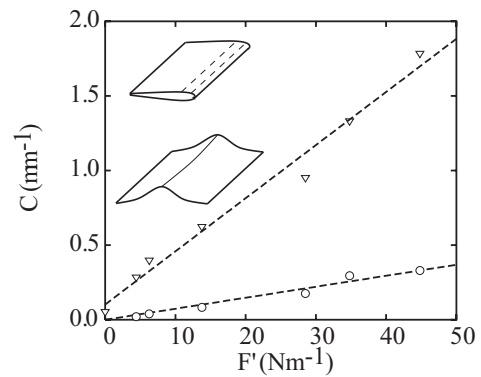}
\end{center}
\caption{The curvature $C$ of a folded sheet plotted versus $F'$, the
  force per unit length. The shapes corresponding to the folded
  $(\triangledown)$ and unfolded $(\circ)$ curvatures are shown in the
  inset. These points give a calibration for the forces that cause the
  deformation in crumpled paper.}
\label{fig:known}
\end{figure}

Before analyzing the entire crumpled surface, it is helpful to examine
a single fold and the forces that cause them. Therefore, we folded a
sheet of paper with thickness $h_p = 0.207$ mm, width $x = 150$ mm and
length $y = 588$ mm as shown in the inset to Fig.~\ref{fig:known}
about the center of the $y$-axis (lengthwise). The two edges of the
folded paper are affixed to the laboratory bench to prevent slipping,
then a known mass (force $F$) is placed across the length of the fold
for uniform compression. After the mass is placed on the fold the
radius of the fold $R_o$ is directly measured.  After approximately 30
seconds of waiting time the mass is removed and the paper is unfolded
and placed in the laser scanning apparatus. The inset to
Fig.~\ref{fig:known} shows schematically the resulting form of the
unfolded sheet which is plastically deformed. A least squares fit to a
circle centered at the apex of the fold yields the radius of
curvature, $R_f$.  It should be noted that the radii are measured on
the outer surface and therefore  include the thickness of the paper
used.  These measurements imply a  lower bound on each measured
radius, $R_{o,f} \ge h_p$ where $h_p$ is the paper thickness.

In Fig.~\ref{fig:known} the initial and final curvatures $C$,
($R_o^{-1},R_f^{-1}$) are plotted versus the force per unit length $F'
= F/y$.   We observe that as expected, the curvature of the fold
increases with an increasing applied force. It is also interesting to
note that although there is always plastic deformation present, the
data is linear over two orders of magnitude in applied force.
Furthermore, the timescales for the application of the applied force
in the hand crumpling and the single folding are equivalent allowing
for direct comparisons of the deformations.

\begin{figure}
\begin{center} 
\includegraphics[width=0.35\textwidth]{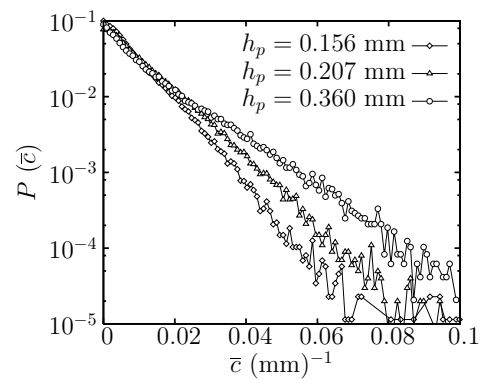}
\end{center}
\vspace{-0.2in}
\caption{The probability distribution of the magnitude of the
curvature $P(\overline{c})$ averaged over regions of $25$ mm$^2$ for
each thickness of paper.  The increase of the curvature for thicker
sheets indicates that the uncrumpling process does not removed the
quenched curvature of the ridges for thicker sheets.}
\label{fig:curves}
\end{figure}

With this understanding of the forces involved to produce a single
fold, we now turn to the analysis of a fully crumpled sheet. One of
the parameters used to assign a label to the point of interest is the
cross-sectional curvature $c$, defined as a plane that  intersects
with the plane of the slope normal, and perpendicular aspect
direction. The distribution of the curvature magnitudes averaged in
boxes of size 5 mm$^2$, $\overline{c}$, are shown in
Fig.~\ref{fig:curves} for all $h_p$.  Note that as $h_p$  increases,
the high curvature tails also increase, indicating that for thicker
sheets, plastic deformation is well preserved after uncrumpling.  This
result is expected by simply considering that the energy required to
unfold a ridge must be proportional to $h_p$.

\begin{figure}
\begin{center} 
\includegraphics[width=0.7\linewidth]{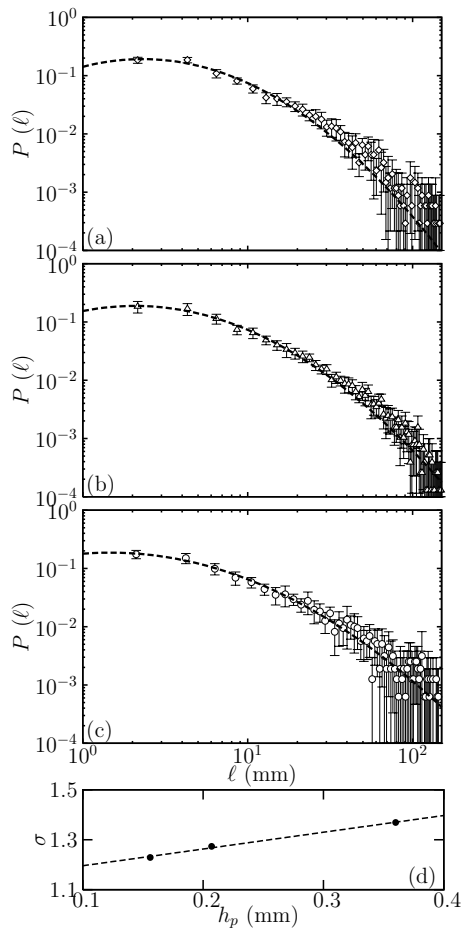}
\end{center}
\caption{The distribution of measured ridge lengths $P(\ell)$ (a) $h_p
  = 0.156$ mm, (b) $h_p = 0.207$ mm, (c) $h_p = 0.360$ mm display.
  The dashed line is a fit to Eq.~\ref{eq:len_dist} and shows good
  agreement and captures the existence of a peak. (d) Width of the
  log-normal distribution $\sigma$ versus the paper thickness $h_p$,
  the dashed line is a linear least squares fit.}
\label{fig:lengths}
\end{figure}

A proposed model of the distribution of energy in a crumpled sheet
states that the process of crumpling forces the majority of the energy
imparted to the sheet will be localized in a network of line-like
ridges~\cite{alex_sci95,alex_pre96,alex_pre97,sethna_n01,wood_physA}.
The structure of the ridge network is said to be that of a simple
hierarchical structure.  The structure is composed of large ridges
that are  randomly and unevenly ``broken''  into smaller ridges that
produces a set of fragmented lengths.  Therefore, a ridge of length
$\ell$ should be   equivalent to the original ridge length times a set
of random variables that correspond to the fraction of ridge left
after each breaking.  The logarithm of the ridge lengths, becomes a
sum of random variables, {\em i.e.}~a random walk.  Therefore,
the probability density function of the ridge length has the form, upon a
change of variables back to lengths~\cite{wood_physA},
\begin{equation}
\frac{dP}{dl} = \frac{1}{\sigma^2\ell} \ e^{-(\log \ell -
 \overline{\log \ell}\ )^2/\sigma^2}.
\label{eq:len_dist}
\end{equation}
A log-normal distribution with a $\ell^{-1}$
pre-factor. To test this prediction, we measured the lengths of the
ridges found in the hand crumpled sheets by identifying the points
that lie along individual ridges via a nearest neighbor criteria.   We
then perform a spline fit to each individual ridge and calculate the
contour length of each spline. In Fig.~\ref{fig:lengths} the
distribution of ridge lengths are plotted for each $h_p$.  The dashed
line is a best fit to Eq.~\ref{eq:len_dist}.  To within experimental
accuracy, the data is well described by this distribution function.

\begin{figure}
\begin{center} 
\includegraphics[width=0.48\linewidth]{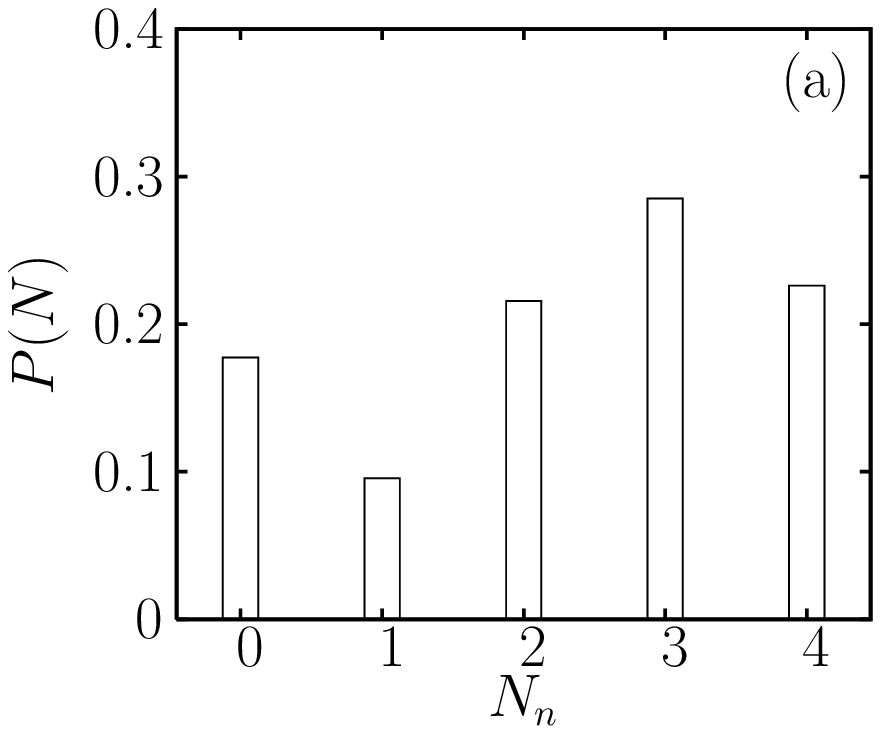}
\includegraphics[width=0.48\linewidth]{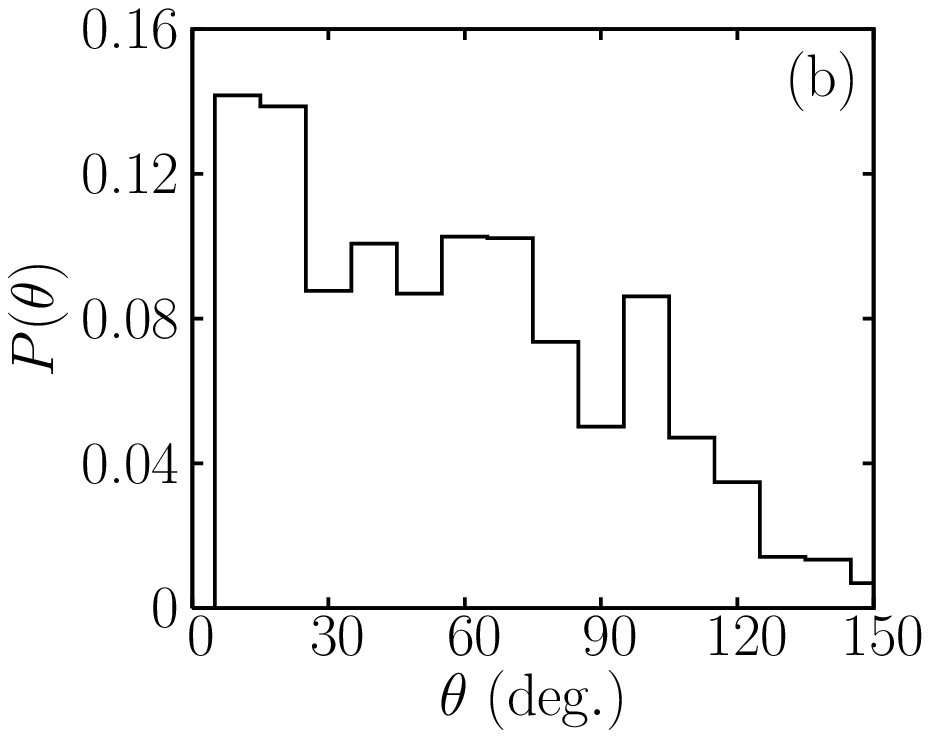}\\
\includegraphics[width=0.48\linewidth]{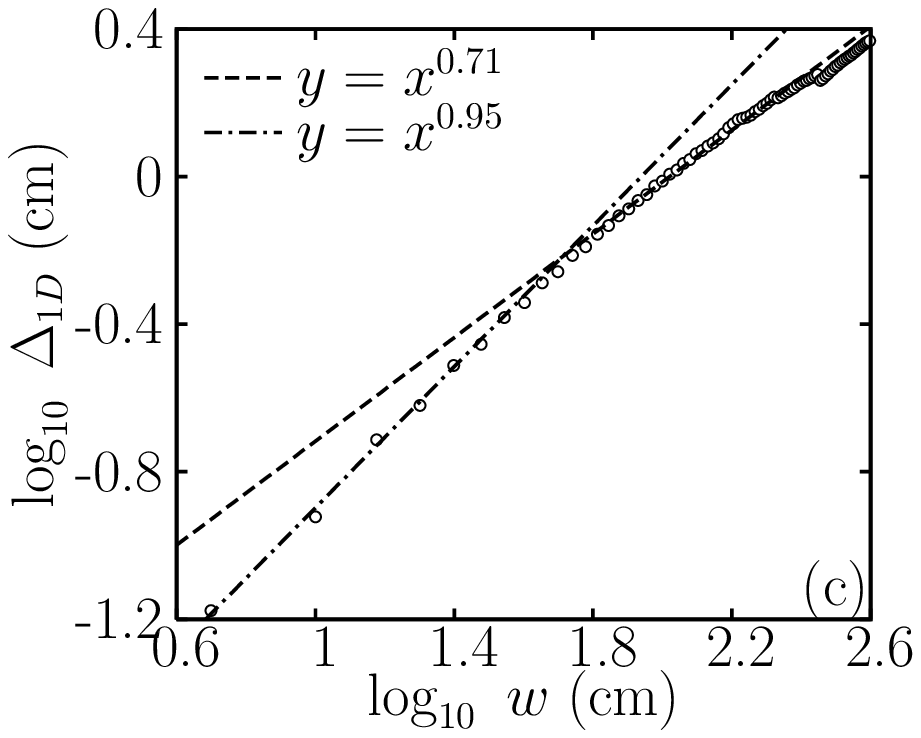}
\includegraphics[width=0.48\linewidth]{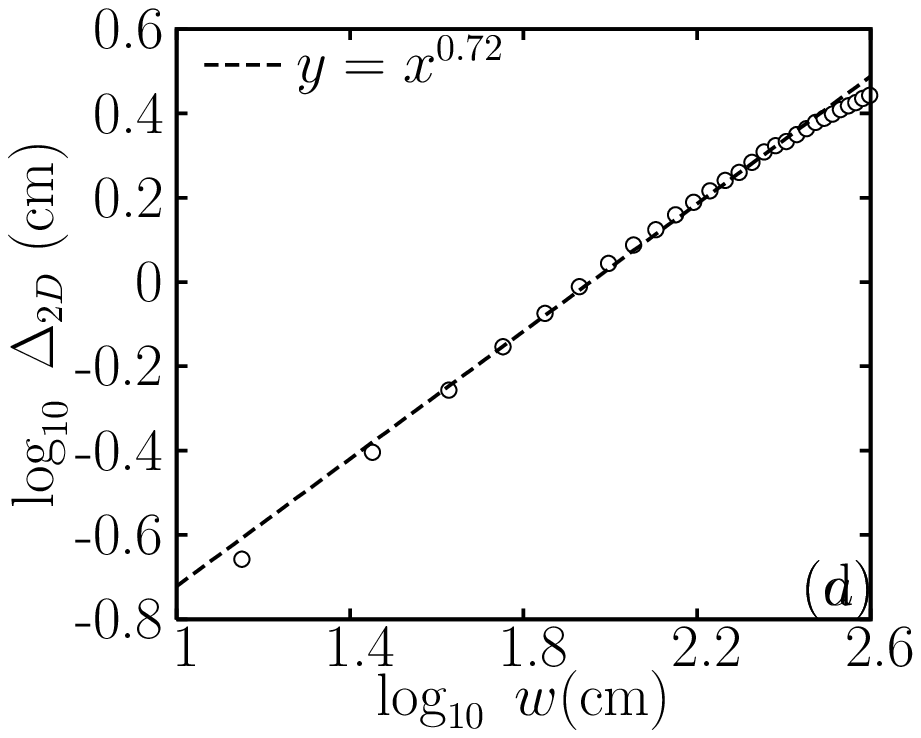}
\end{center}
\caption{(a) Histogram of the number of nearest neighbor ridges at a
  vertex $N_n$.  The high frequency of zero occurrences indicates that
  many ridges dissipate into the paper. (b) Histogram of angle between
  ridges $\theta$ for $N_n=3$. (c) Hurst plot of the one-dimensional
  line crossections of the crumpled paper.  The least square fit gives
  the Hurst exponent $H$.  We observe that for small $w$, there exists
  a scaling usually associated with $1/f$ noise and that for larger
  intervals the scaling follows $H\approx 0.71$. (d) Hurst plot of the
  two-dimensional crumpled gives $H\approx 0.72$. ($h_p =0.207$ mm)} 
\label{fig:hists}
\end{figure}

In theoretical treatments of crumpled surfaces, either
equilibrium~\cite{kantor_86,nelson87}, or 
non-equilibrium~\cite{gomes90,kramer_prl97,di_witten_01}, the
structure of the surface is comprised of randomly oriented ridges that
always intersect at vertices to form a highly connected
network. However, we observe that the {\em connectedness} of the
network is not complete as anticipated. We also measure the number of
neighbors that the ``end'' of a ridge will have, and find that many
ridges do not intersect with other ridges. In Fig.~\ref{fig:hists}(a),
we plot the histogram of the number of nearest neighbors $N_n$, for 20
individual $h_p=0.207$ mm sheets. The existence of ridges with zero
nearest neighbors could have two explanations.  Elastic deformations
do not leave scars of the crumpling process. Therefore, the ridges
that existed during crumpling do not appear when the sheet is
unfolded.  Second, the paper itself may be able to absorb the force
that creates the ridge.  The ridge may essentially dissipate at it's
edges due to the thickness of the paper itself. A question may arise
if the sheet is crumpled ``hard" enough. To answer this question, we
note that squeezing the paper into a smaller $R_c$ than used here
often results in tearing which introduces new complications.   

We also measure the angles that the ridges create with their
neighbors.  In Fig.~\ref{fig:hists}(b) the histogram of the angles
$\theta$, that are created by four ridges is plotted 
({\em i.e.}~ the data in the  $N_n$=3 bin which corresponds to single
d-cones), when they meet at a
vertex.  The values of $\theta$ appear broadly distributed in the
range $0 \ge \theta \ge 180 {\rm  \ deg.}$ 
indicating a somewhat random ordering.  However, a number of
predominant peaks at $\theta \le 20 \ {\rm deg.} \ \theta \approx 60
\ {\rm deg.} \ {\rm
  and} \ \theta \approx 110 \ \rm{deg.}$ are apparent.  The existence
of these peaks indicates that indeed the d-cone geometry dominates the
intersection of ridges when $N_n = 3$ consistent with previous
results~\cite{cerda_prl98,chaieb_prl98}.  However, the precise value
of the predominant opening angle is still debatable.

Finally, to draw connections with previous experimental and
theoretical studies of crumpled surfaces~\cite{roux96}, we also
performed surface scaling measurements.  A self-affine surface
$\zeta(x,y)$, is defined by the scaling of that surface by the
transformation
\begin{equation}
x \to \lambda x,\ y \to \lambda y, \ \zeta \to \lambda^H\zeta,
\label{affine} 
\end{equation} 
where $\lambda$ is an arbitrary scaling factor and $H$ is known as the
Hurst exponent. We measure $H$ directly for each
dimension using the following method.  
Each line scan is sectioned into non-overlapping bins
of width $w$ that are increased in size until $w < L/2$, where $L$ is
the length of the scan.  To ensure that we may consider each scan as
statistically independent, we skip every ten scans.  
 For each bin,
the maximum difference between {\em elevations} within that bin
$\Delta$, is recorded.  This process is continued for each line and
the average of each interval is taken.  For two dimensions the same
process is used.  The width $w$ now represents the length of a box
where the maximum difference is measured.  The relationship,
\begin{equation}
\log_{10}(\Delta)\propto H \log_{10}(w),
\end{equation} 
directly gives the Hurst exponent. The one-dimensional Hurst plot [see
Fig.~\ref{fig:hists}(c)] demonstrates two distinct scaling
regions. For small $w$, $H\sim 0.95$ indicating that for
very short intervals the data seems to be consistent with 
$1/f$ noise.  For larger intervals, the scaling follows $H\sim 0.71$.
The two-dimensional analysis [see Fig.~\ref{fig:hists}(d)] also demonstrates
Hurst scaling. Utilizing one-dimensional scanning profilometry of crumpled 
paper Plourabou\'e and Roux~\cite{roux96} have reported $H = 0.88$.   In a theoretical investigation, based
on a lattice model of crumpled paper, Tzschichholz, {\em et
al.}~\cite{roux95} reported $H=1.0$. Both values are in contrast with our
results.


In summary, we have examined the structure of a crumpled sheet of
paper using laser aided surface topography that allows for the
complete reconstruction of the crumpled surface to unprecedented
precision. While certain features such as ridge length distributions
are found to be consistent with universal scaling laws, many
subtleties are also revealed.  The existence of a linear relationship
between the applied stress and the resultant curvature of the plastic
deformations implies that the surface
curvatures can be mapped to the force distribution of
crumpled surfaces. Due to the linearity of the response, a simpler theory
may be possible to explain our observations.
Additionally, we observe a considerable fraction of ridges end without
bifurcating.  Some of the differences may arise because of the fact
that virtual surfaces which can pass through each were considered in
previous analysis. Including interactions between surfaces as the
paper is squeezed presents new and interesting challenges that needs
to be addressed by further theoretical work.

\begin{acknowledgments}
We thank T. Witten, R. Mukhopadhyay, L. Mahadevan, and A. Boudaoud for
stimulating discussions.
\end{acknowledgments}
 
 \bibliographystyle{apsrev}


\begin{thebibliography}{22}
\expandafter\ifx\csname natexlab\endcsname\relax\def\natexlab#1{#1}\fi
\expandafter\ifx\csname bibnamefont\endcsname\relax
  \def\bibnamefont#1{#1}\fi
\expandafter\ifx\csname bibfnamefont\endcsname\relax
  \def\bibfnamefont#1{#1}\fi
\expandafter\ifx\csname citenamefont\endcsname\relax
  \def\citenamefont#1{#1}\fi
\expandafter\ifx\csname url\endcsname\relax
  \def\url#1{\texttt{#1}}\fi
\expandafter\ifx\csname urlprefix\endcsname\relax\def\urlprefix{URL }\fi
\providecommand{\bibinfo}[2]{#2}
\providecommand{\eprint}[2][]{\url{#2}}

\bibitem[{\citenamefont{Nelson et~al.}(1988)\citenamefont{Nelson, Piran, and
  Weinberg}}]{nelson87}
\bibinfo{author}{\bibfnamefont{D.}~\bibnamefont{Nelson}},
  \bibinfo{author}{\bibfnamefont{T.}~\bibnamefont{Piran}}, \bibnamefont{and}
  \bibinfo{author}{\bibfnamefont{S.}~\bibnamefont{Weinberg}},
  \emph{\bibinfo{title}{Statistical mechanics of membranes and surfaces}}
  (\bibinfo{publisher}{World Scientific, Singapore}, \bibinfo{year}{1988}).

\bibitem[{\citenamefont{Kramer and Witten}(1997)}]{kramer_prl97}
\bibinfo{author}{\bibfnamefont{E.~M.} \bibnamefont{Kramer}} \bibnamefont{and}
  \bibinfo{author}{\bibfnamefont{T.~A.} \bibnamefont{Witten}},
  \bibinfo{journal}{Phys. Rev. Lett.} \textbf{\bibinfo{volume}{78}},
  \bibinfo{pages}{1303} (\bibinfo{year}{1997}).

\bibitem[{\citenamefont{Cerda and Mahadevan}(2003)}]{cerda_prl03}
\bibinfo{author}{\bibfnamefont{E.}~\bibnamefont{Cerda}} \bibnamefont{and}
  \bibinfo{author}{\bibfnamefont{L.}~\bibnamefont{Mahadevan}},
  \bibinfo{journal}{Phys. Rev. Lett.} \textbf{\bibinfo{volume}{90}},
  \bibinfo{pages}{074302} (\bibinfo{year}{2003}).

\bibitem[{\citenamefont{Cerda and Mahadevan}(1998)}]{cerda_prl98}
\bibinfo{author}{\bibfnamefont{E.}~\bibnamefont{Cerda}} \bibnamefont{and}
  \bibinfo{author}{\bibfnamefont{L.}~\bibnamefont{Mahadevan}},
  \bibinfo{journal}{Phys. Rev. Lett.} \textbf{\bibinfo{volume}{80}},
  \bibinfo{pages}{2358} (\bibinfo{year}{1998}).

\bibitem[{\citenamefont{Cerda et~al.}(1999)\citenamefont{Cerda, Chaieb, Melo,
  and Mahadevan}}]{cerda_n99}
\bibinfo{author}{\bibfnamefont{E.}~\bibnamefont{Cerda}},
  \bibinfo{author}{\bibfnamefont{S.}~\bibnamefont{Chaieb}},
  \bibinfo{author}{\bibfnamefont{F.}~\bibnamefont{Melo}}, \bibnamefont{and}
  \bibinfo{author}{\bibfnamefont{L.}~\bibnamefont{Mahadevan}},
  \bibinfo{journal}{Nature} \textbf{\bibinfo{volume}{401}}, \bibinfo{pages}{46}
  (\bibinfo{year}{1999}).

\bibitem[{\citenamefont{Boudaoud et~al.}(2000)\citenamefont{Boudaoud,
  Patr\'icio, Couder, and Amar}}]{arezki_n00}
\bibinfo{author}{\bibfnamefont{A.}~\bibnamefont{Boudaoud}},
  \bibinfo{author}{\bibfnamefont{P.}~\bibnamefont{Patr\'icio}},
  \bibinfo{author}{\bibfnamefont{Y.}~\bibnamefont{Couder}}, \bibnamefont{and}
  \bibinfo{author}{\bibfnamefont{M.~B.} \bibnamefont{Amar}},
  \bibinfo{journal}{Nature} \textbf{\bibinfo{volume}{407}},
  \bibinfo{pages}{718} (\bibinfo{year}{2000}).

\bibitem[{\citenamefont{Cerda et~al.}(2002)\citenamefont{Cerda, Ravi-Shankar,
  and Mahadevan}}]{cerda_n02}
\bibinfo{author}{\bibfnamefont{E.}~\bibnamefont{Cerda}},
  \bibinfo{author}{\bibfnamefont{K.}~\bibnamefont{Ravi-Shankar}},
  \bibnamefont{and}
  \bibinfo{author}{\bibfnamefont{L.}~\bibnamefont{Mahadevan}},
  \bibinfo{journal}{Nature} \textbf{\bibinfo{volume}{53}}, \bibinfo{pages}{579}
  (\bibinfo{year}{2002}).

\bibitem[{\citenamefont{Gomes et~al.}(1989)\citenamefont{Gomes, Jyh, Rodrigues,
  and Furtado}}]{gomes89}
\bibinfo{author}{\bibfnamefont{M.~A.} \bibnamefont{Gomes}},
  \bibinfo{author}{\bibfnamefont{T.~I.} \bibnamefont{Jyh}},
  \bibinfo{author}{\bibfnamefont{I.~M.} \bibnamefont{Rodrigues}},
  \bibnamefont{and} \bibinfo{author}{\bibfnamefont{C.~B.~S.}
  \bibnamefont{Furtado}}, \bibinfo{journal}{J. Phys. D: Appl. Phys.}
  \textbf{\bibinfo{volume}{22}}, \bibinfo{pages}{1217} (\bibinfo{year}{1989}).

\bibitem[{\citenamefont{Kramer and Lobkovsky}(1996)}]{alex_pre96}
\bibinfo{author}{\bibfnamefont{E.~M.} \bibnamefont{Kramer}} \bibnamefont{and}
  \bibinfo{author}{\bibfnamefont{A.~E.} \bibnamefont{Lobkovsky}},
  \bibinfo{journal}{Phys. Rev. E} \textbf{\bibinfo{volume}{53}},
  \bibinfo{pages}{1465} (\bibinfo{year}{1996}).

\bibitem[{\citenamefont{Houle and Sethna}(1996)}]{sethna}
\bibinfo{author}{\bibfnamefont{P.~A.} \bibnamefont{Houle}} \bibnamefont{and}
  \bibinfo{author}{\bibfnamefont{J.~P.} \bibnamefont{Sethna}},
  \bibinfo{journal}{Phys. Rev. E} \textbf{\bibinfo{volume}{54}},
  \bibinfo{pages}{278} (\bibinfo{year}{1996}).

\bibitem[{\citenamefont{Blair}(2004)}]{blair04}
\bibinfo{author}{\bibfnamefont{D.~L.} \bibnamefont{Blair}}, Ph.D. thesis,
  \bibinfo{school}{Clark University} (\bibinfo{year}{2004}).

\bibitem[{\citenamefont{Wood}(1996)}]{wood}
\bibinfo{author}{\bibfnamefont{J.}~\bibnamefont{Wood}}, Ph.D. thesis,
  \bibinfo{school}{University of Leicester} (\bibinfo{year}{1996}).

\bibitem[{\citenamefont{Lobkovsky et~al.}(1995)\citenamefont{Lobkovsky,
  Gentges, Li, Morse, and Witten}}]{alex_sci95}
\bibinfo{author}{\bibfnamefont{A.~E.} \bibnamefont{Lobkovsky}},
  \bibinfo{author}{\bibfnamefont{S.}~\bibnamefont{Gentges}},
  \bibinfo{author}{\bibfnamefont{H.}~\bibnamefont{Li}},
  \bibinfo{author}{\bibfnamefont{D.}~\bibnamefont{Morse}}, \bibnamefont{and}
  \bibinfo{author}{\bibfnamefont{T.~A.} \bibnamefont{Witten}},
  \bibinfo{journal}{Science} \textbf{\bibinfo{volume}{270}},
  \bibinfo{pages}{1482} (\bibinfo{year}{1995}).

\bibitem[{\citenamefont{Lobkovsky and Witten}(1997)}]{alex_pre97}
\bibinfo{author}{\bibfnamefont{A.~E.} \bibnamefont{Lobkovsky}}
  \bibnamefont{and} \bibinfo{author}{\bibfnamefont{T.~A.}
  \bibnamefont{Witten}}, \bibinfo{journal}{Phys. Rev. E}
  \textbf{\bibinfo{volume}{55}}, \bibinfo{pages}{1577} (\bibinfo{year}{1997}).

\bibitem[{\citenamefont{Sethna et~al.}(2001)\citenamefont{Sethna, Dahmen, and
  Meyers}}]{sethna_n01}
\bibinfo{author}{\bibfnamefont{J.~P.} \bibnamefont{Sethna}},
  \bibinfo{author}{\bibfnamefont{K.~A.} \bibnamefont{Dahmen}},
  \bibnamefont{and} \bibinfo{author}{\bibfnamefont{C.~R.}
  \bibnamefont{Meyers}}, \bibinfo{journal}{Nature}
  \textbf{\bibinfo{volume}{410}}, \bibinfo{pages}{242} (\bibinfo{year}{2001}).

\bibitem[{\citenamefont{Wood}(2002)}]{wood_physA}
\bibinfo{author}{\bibfnamefont{A.~J.} \bibnamefont{Wood}},
  \bibinfo{journal}{Physica A} \textbf{\bibinfo{volume}{313}},
  \bibinfo{pages}{83} (\bibinfo{year}{2002}).

\bibitem[{\citenamefont{Kantor et~al.}(1986)\citenamefont{Kantor, Kardar, and
  Nelson}}]{kantor_86}
\bibinfo{author}{\bibfnamefont{Y.}~\bibnamefont{Kantor}},
  \bibinfo{author}{\bibfnamefont{M.}~\bibnamefont{Kardar}}, \bibnamefont{and}
  \bibinfo{author}{\bibfnamefont{D.~R.} \bibnamefont{Nelson}},
  \bibinfo{journal}{Phys. Rev. Lett.} \textbf{\bibinfo{volume}{57}},
  \bibinfo{pages}{791} (\bibinfo{year}{1986}).

\bibitem[{\citenamefont{{Gomes} et~al.}(1990)\citenamefont{{Gomes}, {Jyh}, and
  {Ren}}}]{gomes90}
\bibinfo{author}{\bibfnamefont{M.~A.~F.} \bibnamefont{{Gomes}}},
  \bibinfo{author}{\bibfnamefont{T.~I.} \bibnamefont{{Jyh}}}, \bibnamefont{and}
  \bibinfo{author}{\bibfnamefont{T.~I.} \bibnamefont{{Ren}}},
  \bibinfo{journal}{Journal of Physics A Mathematical General}
  \textbf{\bibinfo{volume}{23}}, \bibinfo{pages}{L1281} (\bibinfo{year}{1990}).

\bibitem[{\citenamefont{DiDonna and Witten}(2001)}]{di_witten_01}
\bibinfo{author}{\bibfnamefont{B.~A.} \bibnamefont{DiDonna}} \bibnamefont{and}
  \bibinfo{author}{\bibfnamefont{T.~A.} \bibnamefont{Witten}},
  \bibinfo{journal}{Phys. Rev. Lett.} \textbf{\bibinfo{volume}{87}},
  \bibinfo{pages}{206105} (\bibinfo{year}{2001}).

\bibitem[{\citenamefont{Cha\"ieb et~al.}(1998)\citenamefont{Cha\"ieb, Melo, and
  G\'eminard}}]{chaieb_prl98}
\bibinfo{author}{\bibfnamefont{S.}~\bibnamefont{Cha\"ieb}},
  \bibinfo{author}{\bibfnamefont{F.}~\bibnamefont{Melo}}, \bibnamefont{and}
  \bibinfo{author}{\bibfnamefont{J.-C.} \bibnamefont{G\'eminard}},
  \bibinfo{journal}{Phys. Rev. Lett.} \textbf{\bibinfo{volume}{80}},
  \bibinfo{pages}{2354} (\bibinfo{year}{1998}).

\bibitem[{\citenamefont{Plourabou\'e and Roux}(1996)}]{roux96}
\bibinfo{author}{\bibfnamefont{F.}~\bibnamefont{Plourabou\'e}}
  \bibnamefont{and} \bibinfo{author}{\bibfnamefont{S.}~\bibnamefont{Roux}},
  \bibinfo{journal}{Physica A} \textbf{\bibinfo{volume}{227}},
  \bibinfo{pages}{173} (\bibinfo{year}{1996}).

\bibitem[{\citenamefont{Tzschichholz et~al.}()\citenamefont{Tzschichholz,
  Hansen, and Roux}}]{roux95}
\bibinfo{author}{\bibfnamefont{F.}~\bibnamefont{Tzschichholz}},
  \bibinfo{author}{\bibfnamefont{A.}~\bibnamefont{Hansen}}, \bibnamefont{and}
  \bibinfo{author}{\bibfnamefont{S.}~\bibnamefont{Roux}},
  \bibinfo{note}{cond-mat/9507056}.

\end{thebibliography}

\end{document}